\documentclass{article}
\usepackage{spconf,amsmath,graphicx}
\usepackage{cite}
\usepackage{hyperref}
\usepackage{cleveref}
\usepackage{multirow}
\usepackage{amssymb}
\usepackage{xfrac}
\usepackage{booktabs}
\usepackage[dvipsnames]{xcolor}
\usepackage{float}

\title{SWIM: Short-Window CNN Integrated with Mamba for\\ EEG-Based Auditory Spatial Attention Decoding}
\name{Ziyang Zhang$^1$, Andrew Thwaites$^2$, Alexandra Woolgar$^3$, Brian Moore$^3$, Chao Zhang$^1$\sthanks{Corresponding author.}\vspace{-0.4cm}}
\address{$^1$ Tsinghua University, China\\
$^2$University College London, United Kingdom \\
$^3$University of Cambridge, United Kingdom \\
\texttt{ziyang-z24@mails.tsinghua.edu.cn, cz277@tsinghua.edu.cn}}
\ninept

\begin{document}

\maketitle

\begin{abstract}
    In complex auditory environments, the human auditory system possesses the remarkable ability to focus on a specific speaker while disregarding others. In this study, a new model named SWIM, a short-window convolution neural network (CNN) integrated with Mamba, is proposed for identifying the locus of auditory attention (left or right) from electroencephalography (EEG) signals without relying on speech envelopes.
    SWIM consists of two parts. The first is a short-window CNN (SW$_\text{CNN}$), which acts as a short-term EEG feature extractor and achieves a final accuracy of 84.9\% in the leave-one-speaker-out setup on the widely used KUL dataset. This improvement is due to the use of an improved CNN structure, data augmentation, multitask training, and model combination.
    The second part, Mamba, is a sequence model first applied to auditory spatial attention decoding to leverage the long-term dependency from previous SW$_\text{CNN}$ time steps. By joint training SW$_\text{CNN}$ and Mamba, the proposed SWIM structure uses both short-term and long-term information and achieves an accuracy of 86.2\%, which reduces the classification errors by a relative 31.0\% compared to the previous state-of-the-art result. The source code is available at \texttt{\url{https://github.com/windowso/SWIM-ASAD}}.
\end{abstract}


\section{Introduction}

Humans have a remarkable ability for selective hearing. Given complex auditory environments with multiple speakers, multiple conversations and overlapping speech, the human auditory system can selectively ignore and filter out the speech of other speakers by focusing on the specific speaker of interest, a phenomenon known as the cocktail party effect\cite{cherry1953some}. However, this auditory task proves challenging for individuals with hearing impairments. The reduced spectral and temporal resolution of the early auditory system associated with hearing loss adversely affects cocktail party processing and hearing aids have a limited ability to overcome these deficits. A possible solution would be to enhance the voice of the attended speaker, but first they must be identified. One way could be to identify the attended speaker via eye gaze or gestures, but these methods are not sufficiently robust. A more effective solution could involve using the listener's neural activity, as measured through EEG or other devices, to directly determine the attended speaker, a process known as auditory attention decoding. 

Previous works in auditory attention decoding have primarily studied stimulus reconstruction, where post-stimulus brain activity is used to decode and reconstruct the envelope of the attended speech stimulus \cite{de2020machine, o2015attentional, geirnaert2021electroencephalography}. 
More recent studies focused on the direct decoding of the spatial location of the attended speaker
\cite{vandecappelle2021eeg, geirnaert2020fast, cai2023bio, cai2023brain, jiang2022detecting, xu2023densenet, zhang2020eeg, zhang2023learnable, mahjoory2023convolutional, su2022stanet, pahuja2023xanet}, a paradigm known as auditory spatial attention decoding (ASAD). ASAD eliminates the need for a clean speech stimulus, thus advancing the development of practical neuro-steered hearing prostheses. 
To develop high-performance ASAD for neuro-steered hearing, the models used need to satisfy the following abilities: 
1) The ability to detect and respond to the change in the user's auditory attention rapidly in a short time window;
2) the ability to use the features embedded in the long-term EEG signals to improve the detection of the auditory attention direction in the current window.
These abilities help the model to achieve not only stable detection of the listener's long-term auditory attention focus on the same speaker, but also accurate and immediate detection of attention changes when the listener intends to focus on a different speaker. 
%
In the current existing body of literature, convolutional neural networks (CNNs) are applied to EEG signals to achieve high-performance short-term auditory attention decoding for low-latency ASAD \cite{vandecappelle2021eeg}. Transformer-based models including STAnet\cite{su2022stanet} and XAnet\cite{pahuja2023xanet}, are developed to use the attention mechanisms to model long-term temporal information in EEG signals. 
By combining these two structures, it is possible to achieve both long-term stable detection with short-term rapid response abilities using a single model.

In this paper, we propose a novel model structure \textbf{s}hort-\textbf{w}indow CNN \textbf{i}ntegrated with \textbf{M}amba\cite{mamba} (SWIM) that can model both short-term and long-term EEG signal patterns by stacking CNN and Mamba for ASAD. The CNN acts as a feature extraction from raw EEG signals, enabling the acquisition of highly efficient and accurate local spatial-temporal patterns for auditory attention within the current time step.  
The Mamba sequence model uses the features extracted by the CNN as the input features at each time step, incorporating long-term temporal information carried from previous time steps to assist in judging the auditory attention of the current time step. 
Compared to other commonly used sequence models including recurrent neural network (RNN) and Transformer, Mamba which is improved based on the state space model (SSM), combines the advantages of high inference efficiency from RNN and the high training efficiency from Transformer, making it more suitable for the deployment to cost-sensitive neuro-steered hearing devices.

Specifically, the short-window CNN (SW$_\text{CNN}$) uses a short CNN kernel time window to focus on local patterns of the EEG signals within each 1-second (s) window. Window overlapping and time masking data augmentation methods along with multitask training are used to improve SW$_\text{CNN}$. Further by combining two SW$_\text{CNN}$ models trained with different input channel configurations, an accuracy of 84.9\% is achieved in the Leave-one-speaker-out setup of the widely used KUL dataset\cite{biesmans2016auditory}, which received a 4.9\% absolute increase in accuracy over the previous state-of-the-art (SOTA) result using the same setup. 
The Mamba model takes a sequence of SW$_\text{CNN}$ outputs at a time step shift of 0.125s, which not only enables SWIM to leverage long-term temporal dependency across time steps but also enables a rapid response to auditory attention change with a latency of only 0.125s. Although Mamba has been applied to EEG signal processing \cite{panchavatimentality, zhanga2024mssc, lee2024neuronet, zhou2024benchmarking}, this work is the first to use it for ASAD, and this is also the first to explore the stacked SW$_\text{CNN}$ and Mamba structure, similar to the previous prevalent structures by stacking CNN with RNN or Transformer \cite{sainath2015convolutional,schneider2019wav2vec}. Experimental results showed that SWIM further improved the Leave-one-speaker-out accuracy to 86.2\%. Comprehensive analyses are provided to understand the importance of different EEG channels and insights into different setups \cite{rotaru2024we}.

\section{Related Work}

\textbf{ASAD Models.} The ASAD task does not require clean speech envelopes as input, making it suitable for practical applications in neuro-steered hearing devices. Numerous different models have been proposed for ASAD, such as traditional methods like common spatial patterns (CSP) \cite{geirnaert2020fast}. Graph neural networks \cite{scarselli2008graph} use EEG graph transformed from EEG data for leveraging the positioning information of EEG channels \cite{cai2023brain, jiang2022detecting}. The neuroscience-inspired spiking neural network offers a fast, accurate, and energy-efficient ASAD \cite{cai2023bio}. DenseNet-3D converts EEG data into a three-dimensional (3D) arrangement containing spatial-temporal information\cite{xu2023densenet}. Transformer-based models including STAnet\cite{su2022stanet} and XAnet\cite{pahuja2023xanet} use attention mechanism in decoding. CNNs are model architectures that have shown good performance and efficiency in extracting features from EEG for ASAD tasks\cite{vandecappelle2021eeg}.

\noindent \textbf{Sequence Models.} Sequence models are typically used to model dependencies in time series data. RNN, as the earliest proposed sequence model, capture historical information through hidden states, and their variants \cite{hochreiter1997long} are widely used in many sequence modeling tasks. Transformer \cite{vaswani2017attention}, the recent prevalent sequence model, use the attention mechanism to handle the temporal dependencies, addressing RNN's issues with parallel training across time and long sequence memory loss. Transformers are highly scalable and have led to the development of applications like large language models. Mamba \cite{mamba, mamba2}, a recently proposed sequence model of prevalence, is an improvement based on SSM. Mamba combines the benefits from both Transformer and RNN, which can be trained by parallelized across time and is good at modeling long sequences like Transformer, while performing computational and memory-efficient streaming inference like RNN. 
Overall, Mamba combines the fast autoregressive inference of RNN with the efficient parallel training of the Transformer, balancing performance and efficiency.

\noindent \textbf{EEG Data Augmentation and Multitask Training.} Data augmentation and multitask training are commonly used in deep learning. Given that ASAD datasets are typically of limited sizes, often comprising only dozens to tens of hours of data, the application of data augmentation and multitask training becomes crucial for making full use of EEG data and labels. 
Many methods have been proposed in the realm of EEG data augmentation \cite{rommel2022data, lashgari2020data, salama2018eeg, yin2017cross, piplani2018faking, panwar2019generating, manor2015convolutional, thodoroff2016learning, zhang2019novel, said2017multimodal}. The approaches typically involve exploring different overlapping between sliding windows \cite{ali2022enhancing, kwak2017convolutional}, and time masking that sets a portion of the signal to zero\cite{mohsenvand2020contrastive}. 
There are also data augmentation methods operating in the frequency domain \cite{mohsenvand2020contrastive} and in channel dimension \cite{saeed2021learning, krell2017rotational}. With regards to multitasking, reconstruction of the speech envelope has been considered as an auxiliary task to improve ASAD accuracy\cite{zhang2020eeg}.

\section{Methods}

SWIM consists of two parts: a short-window CNN (SW$_\text{CNN}$) and a Mamba sequence model. SW$_\text{CNN}$ acts as an EEG feature extractor, responding rapidly and reliably within a short window. Mamba captures the temporal dependencies across different time steps over long EEG sequences, using the information from previous time steps to improve the prediction of the current time step. Additionally, EEG data augmentation techniques are applied to both SW$_\text{CNN}$ and SWIM. The following sections will introduce SW$_\text{CNN}$, Mamba, SWIM and EEG data augmentation techniques.

\subsection{Short-Window CNN}
\begin{figure}
    \centering
    \includegraphics[width=\linewidth]{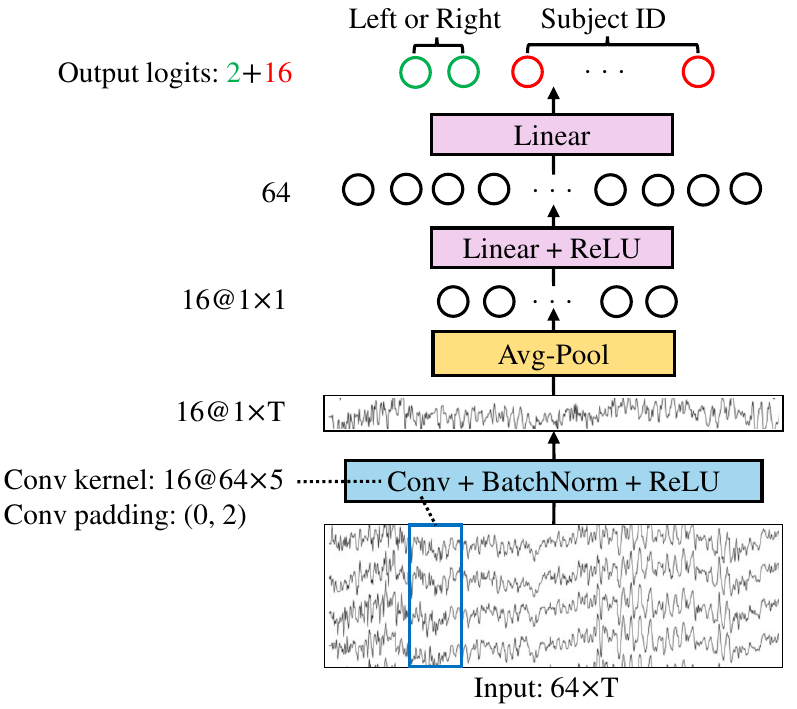}
    \vspace{-0.3cm}
    \caption{The architecture of SW$_\text{CNN}$. The input is a decision window of an EEG signal with 64 channels and $T$ samples. The output is logits for classifying attention location and subject ID. In this figure, the number in front of the @ represents the model channel dimension, and the + represents the vector concatenation of two dimensions.}
    \vspace{-0.3cm}
    \label{fig:cnn}
\end{figure}

The SW$_\text{CNN}$ in Fig.~\ref{fig:cnn} is improved based on the CNN structure used in  \cite{vandecappelle2021eeg}. The major structural differences are: 1) the kernel time window is reduced from 17 to 5; 2) the channels of CNN are increased from 5 to 16, and the size of the fully connected (FC) layer is increased from 5 to 64 and batch normalization (BatchNorm) is added. 

The EEG signal is represented as a matrix in \(\mathbb{R}^{64\times T}\), indicating the presence of 64 channels and \(T\)-length decision window. The initial layer within SW$_\text{CNN}$ incorporates a CNN layer characterized by a kernel dimension of \(64\times 5\), a padding schema of \(\left(0, 2\right)\), and an output channel count of \(16\). This layer is succeeded by using BatchNorm and a ReLU activation function. Consequently, the output generated by this CNN layer is structured as a \(\mathbb{R}^{16\times 1\times T}\) matrix.
After the CNN layer, an average pooling operation is employed, which performs an averaged pooling across time. This operation effectively condenses each time series into a singular value, resulting in an output dimensionality of \(\mathbb{R}^{16\times 1\times 1}\). Following the pooling operation, SW$_\text{CNN}$ incorporates two FC layers. The initial FC layer accepts an input in \(\mathbb{R}^{16}\), projects this input into a \(\mathbb{R}^{64}\) space, and is further processed by a ReLU activation function. The subsequent FC layer, in turn, processes the \(\mathbb{R}^{64}\) input to produce an output in \(\mathbb{R}^{18}\).

SW$_\text{CNN}$ is trained on EEG data from multiple subjects, enabling it to simultaneously classify the direction of auditory spatial attention and identify the originating subject. Within the output layer, the two-dimensional (-dim) output layer is associated with the classification of auditory spatial attention direction, and the 16-dim output layer is associated with the classification of the 16 subject IDs.
Let \(\mathcal{L}_\text{locus}\) and \(\mathcal{L}_\text{subject}\) are the cross-entropy loss functions computed based on the spatial attention locus and the subject ID correspondingly. The final loss function, \(\mathcal{L}_\text{total}\), is computed as follows:
\begin{equation}
\mathcal{L}_\text{total}=\mathcal{L}_\text{locus}+\gamma\cdot\mathcal{L}_\text{subject}.
\label{eq:1}
\end{equation}
Eqn.~\eqref{eq:1} combines the primary and auxiliary loss, with \(\gamma\) being a scalar to scale the auxiliary loss. The use of the auxiliary subject loss enhances the model's capacity to extract subject-dependent features by making it learn diverse subjects.

\subsection{The Mamba Model}
Mamba is a sequence model designed to achieve highly efficient and accurate long sequence modeling. Mamba is developed based on SSM\cite{gu2022efficiently, pmlr-v162-goel22a}, which can be represented as follows
\begin{align}
\mathbf{h}_n&= \overline{\mathbf{A}}\,\mathbf{h}_{n-1} + \overline{\mathbf{B}}\,\mathbf{x}_n\\
\mathbf{y}_n&=\mathbf{C}\,\mathbf{h}_n.
\end{align}
By compressing history information into the state \( \mathbf{h}_n \), SSM can handle very long sequences while maintaining low computational and memory requirements. Due to the good properties of SSM, it has efficient parallel training like the Transformer and fast auto-regressive generation like the RNN during inference. 

Mamba introduces two key improvements to SSM. First, Mamba incorporates a selection mechanism related to the input-to-state matrix, making the information filtering mechanism of SSM input-dependent, achieving comparable effectiveness to the Transformer, which is called Selective SSM (S6). Second, Mamba optimizes the S6 algorithm on hardware, achieving a theoretical time complexity of \( \mathcal{O}(n) \) and space complexity of \( \mathcal{O}(1) \) during training, while the Transformer's theoretical time complexity is \( \mathcal{O}(n^2) \) and space complexity is \( \mathcal{O}(n) \), thus Mamba is computationally more efficient than Transformer. 

Mamba is highly suitable for our target scenario: A practical EEG-assisted hearing aid device should respond to the change of the user's auditory attention rapidly in low latency while being able to leverage long-term patterns embedded in the previous time steps to aid such short-window judgments. The history length could span hours or even days, placing high demands on the model's inference time complexity. 
Compared to Transformer, Mamba’s inference time complexity is \( \mathcal{O}(1) \) instead of \( \mathcal{O}(n) \), which is more suitable for streaming ASAD in practical scenarios. Additionally, hearing aid devices often have limited memory, and Mamba’s low storage complexity of \( \mathcal{O}(1) \) is better suited for edge deployment compared to the Transformer's storage complexity of \( \mathcal{O}(n) \).

\subsection{From Short Window to Long Sequence}

While SW$_\text{CNN}$ can perform accurate and efficient ASAD based on short windows, it does not use the long-term information embedded in the previous time steps of the EEG signals, which limits its performance upper bound when making more rapid detection of focus changes with a small time window shift. 
In contrast, Mamba can achieve highly accurate long sequence modeling by leveraging the temporal dependencies to enhance the performance of the decision of the current time step. Therefore, it is natural and beneficial to extend from a short window to a long sequence. To achieve this, SWIM, a novel structure combining SW$_\text{CNN}$ and Mamba is proposed for ASAD. This approach retains SW$_\text{CNN}$'s advantage of rapid response in based on local patterns within a short window while further improving the accuracy and latency by leveraging the temporal dependency from the entire EEG signals.

\begin{figure}[h]
    \centering
    \includegraphics[width=1\linewidth]{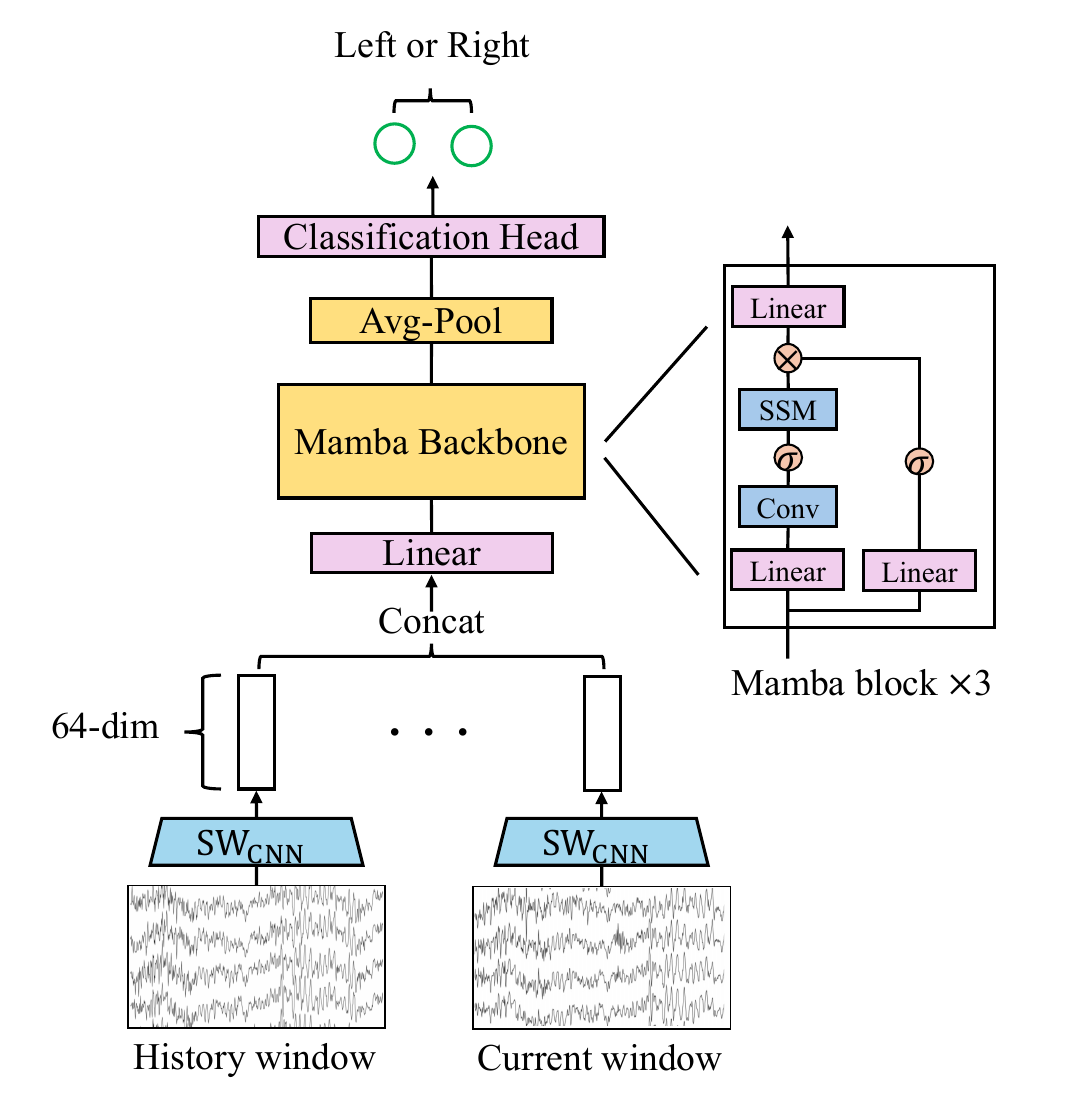}
    \vspace{-0.3cm}
    \caption{The architecture of SWIM. The SW$_\text{CNN}$ is shown in Fig.~\ref{fig:cnn} with the classification head removed, so the output of SW$_\text{CNN}$ is a 64-dim hidden feature. The hidden features from history windows are concatenated with it from the current window as input of Mamba. Then Mamba utilize this input to classify the auditory attention direction of the current window. In this figure, \(\times\) means multiplication and \(\sigma\) means an activation in the Mamba block.}
    \vspace{-0.3cm}
    \label{fig:SWIM}
\end{figure}

As shown in Fig.~\ref{fig:SWIM}, SW$_\text{CNN}$ with classification head removed extracts 64-dim hidden features from the 64 channels and $T$-length decision window of raw EEG signals. Mamba does not directly act on raw EEG data. Instead, it uses features extracted by SW$_\text{CNN}$ as input. The hidden features extracted from the current window are concatenated with those extracted from previous CNN decision windows along the time dimension to form a matrix in \( \mathbb{R}^{64 \times N} \). This matrix is used as Mamba's input, and Mamba outputs a binary classification result for the current window, indicating the direction of auditory spatial attention (left or right).

The Mamba model structure consists of three parts. First, the Mamba backbone network has 3 layers of Mamba blocks, each with a model dimension of 64 and a state dimension of 16. Second, there is an FC layer between the output from SW$_\text{CNN}$ and the Mamba backbone network, with an input dimension of 64 and an output dimension matching the Mamba model dimension. Third, following the Mamba backbone network, there is an average pooling layer and a classification head. The output size of the Mamba backbone network matches the input size, resulting in an output matrix of \( \mathbb{R}^{64 \times N} \). For classification, the average pooling layer reduces the output to a 64-dimensional vector, which is then processed by the classification head, an FC layer that converts the 64-dimensional vector to a 2-dimensional output for classifying auditory attention direction.

\subsection{EEG Data Augmentation Techniques}

\begin{figure}
    \centering
    \includegraphics[width=1\linewidth]{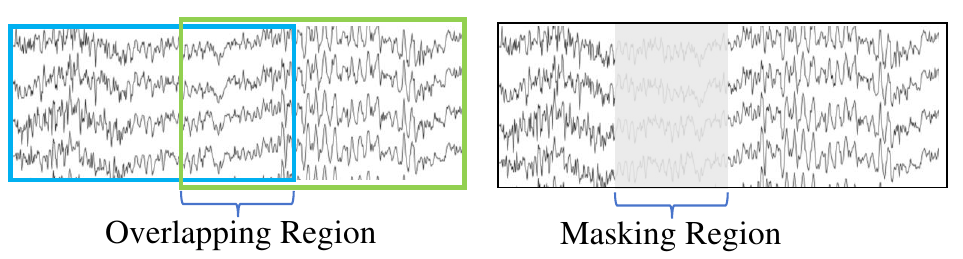}
    \vspace{-0.3cm}
    \caption{In the left figure, the two boxes represent two decision windows. If the overlapping ratio is not zero, then they will have overlapping regions. In the right figure, the masking region in a decision window will be set to zero.}
    \vspace{-0.3cm}
    \label{fig:overlap mask demo}
\end{figure}

Two data augmentation techniques are proposed to act on the EEG matrix directly, to facilitate the model's exposure to more diverse EEG data segments and durations.
\begin{enumerate}
    \item \textbf{Overlapping}: EEG data is segmented into many overlapped decision windows as shown in Fig.~\ref{fig:overlap mask demo}, thereby increasing the diversity of the data samples available for training.
    
    \item \textbf{Time Masking}: The time masking is
conducted by SpecAugmentation\cite{park2019specaugment}. \(t\) consecutive time steps \(\left[t_0, t_0 + t\right)\) are masked as shown in Fig.~\ref{fig:overlap mask demo}. \(t\) is randomly drawn from a uniform distribution ranging from 0 to a predefined time mask parameter \(\tau\), ensuring variability in the extent of masking applied across different instances. Subsequently, the starting point \(t_0\) is chosen from the interval \(\left[0, T - t\right)\).
\end{enumerate}

\section{Experimental Setup}

\subsection{Data and Preprocessing}

Our experiments used the widely used KUL dataset \cite{biesmans2016auditory}, which comprises EEG recordings from 16 subjects with normal hearing. Each subject underwent 8 trials, with each trial lasting 6 minutes. Subjects were instructed to focus on one of two competing speakers that were simultaneously active, positioned at \(\pm 90^{\circ}\) in the azimuth direction, representing the left and right locations respectively. The dataset features four stories delivered by three speakers: Speaker1 and Speaker2 narrate Story1 and Story2 respectively, while Speaker3 tells Story3 and Story4. The dataset was recorded using a 64-channel BioSemi ActiveTwo EEG system, followed by pre-processing with down sampling to 128Hz and bandpass filtering. At last, the data is normalized to per-window-based zero mean and per-trial-based unit standard deviation for each EEG channel.

\subsection{Evaluation Approach}

The dataset is partitioned according to three strategies:

\begin{enumerate}
    \item \textbf{Every-trial}: The evaluation approach ensures that the training, validation, and test datasets encompass data from Every-trial. Specifically, the initial 70\% of data from each trial is allocated to the training set, the subsequent 15\% to the validation set, and the final 15\% to the test set. 
    
    \item \textbf{Leave-one-speaker-out}: The dataset consists of data from three speakers. Speaker1 and Speaker2 each narrated a single story, and Speaker3 narrated two stories. To ensure there is sufficient training data, the strategy will focus on leaving Speaker1 and Speaker2 out. For example, Leave-Speaker1-out designates the trials as test data if its subject paid attention to Speaker1's narrations, while the remaining trials are divided into training and validation sets with an 85/15\% split. This method is replicated for Speaker2, culminating in the outcome being the mean of the results obtained from the ``leave-Speaker1-out" and ``leave-Speaker2-out". 
    
    \item \textbf{Leave-one-subject-out}: Designates all trials from a single subject as test data, with the remainder of the data being split into training and validation sets according to a ratio of 85/15\%. The outcome is the mean of the results obtained from the ``leave-Subject1-out" to ``leave-Subject16-out".
\end{enumerate}

\subsection{Model and Training Implementation}

SWIM was trained using the PyTorch framework on an NVIDIA RTX 4090 GPU. The Adam optimizer \cite{kingma2014adam} was used and the Cosine Annealing Learning Rate Scheduler was used to adjust the learning rate over the training epochs. Regularization approaches including early stopping based on validation accuracy and weight decay are used to prevent overfitting. Specifically, SW$_\text{CNN}$ was trained with a batch size of 64, initial learning rate of 1e-3, maximum epoch number of 100 and weight decay of 1e-3. SWIM was trained with a batch size of 32, maximum epoch number of 5 and weight decay of 0. When training SWIM, the parameters of SW$_\text{CNN}$ are loaded from a pre-trained version and fine-tuned with an initial learning rate of 1e-5. The Mamba part starts with an initial learning rate of 1e-3. When training $\text{SW}_{\text{CNN}}$ and SWIM, the learning rate was varied from 1e-5 to 1e-2, the batch size was varied from 32 to 128, and the weight decay was varied from 0 to 1e-1. The hyperparameters were selected according to the results of the validation set.

\section{Results and Analysis}

In this section, the results of three experimental setups are reported.
If without explicit specification, the results are derived using a Leave-one-speaker-out setup with a decision window length of 1 second. All experiments are repeated three times with different random seeds, and the mean of the results is reported.

\subsection{Improvement from SW$_\text{CNN}$}

\begin{table}[]
\centering
\caption{SW$_\text{CNN}$ is changed from CNN\cite{vandecappelle2021eeg} through two steps. Step1: Reduce the length of the kernel time window. Step2: Increase the number of channels and add BatchNorm.}
\resizebox{\linewidth}{!}{
\begin{tabular}{@{}ll@{}}
\toprule
Model        & \%Accuracy \\ \midrule
CNN\cite{vandecappelle2021eeg} & 75.2       \\
kernel time window \(17\to 5\)                       & 77.6 (2.4\(\uparrow\)) \\
CNN channels \(5\to 16\) + BatchNorm(SW$_\text{CNN}$) & 80.9 (3.3\(\uparrow\)) \\ \bottomrule
\end{tabular}}
\label{tab: model}
\end{table}

SW$_\text{CNN}$ has a shorter kernel time window and more channels than the structure in \cite{vandecappelle2021eeg} and adds multitask training. 
Without using data augmentation, multitask training or model combination, an ablation study was provided in Table~\ref{tab: model}, showing that the improvement comes from the reduced kernel time window and the increased number of CNN channels. The results found that the use of a long kernel time window with Avg-Pool is inferior to the short kernel time window. Conversely, the increase of CNN channels allows SW$_\text{CNN}$ to capture more EEG patterns and improve ASAD accuracy.

After confirming the structure of SW$_\text{CNN}$, multi-task training was further added, whose results are illustrated in Fig.~\ref{fig:overlap mask multitask}(c). 
With the optimal auxiliary loss weighting factor $\gamma=0.05$, the ASAD decoding accuracy reaches the peak.
When $\gamma$ decreases, the multitask training gradually becomes ineffective and the ASAD accuracy drops. 
When $\gamma$ becomes overly large, for instance, 0.1 or 0.2, the auxiliary loss starts to negatively impact the primary loss, leading to a decrease in accuracy.

\subsection{Data Augmentation on SW$_\text{CNN}$}

Two data augmentation methods, overlapping and time masking, were applied to SW$_\text{CNN}$. The following will explore the improvements brought by each method.

\begin{figure}[]
    \centering
    \includegraphics[width=1\linewidth]{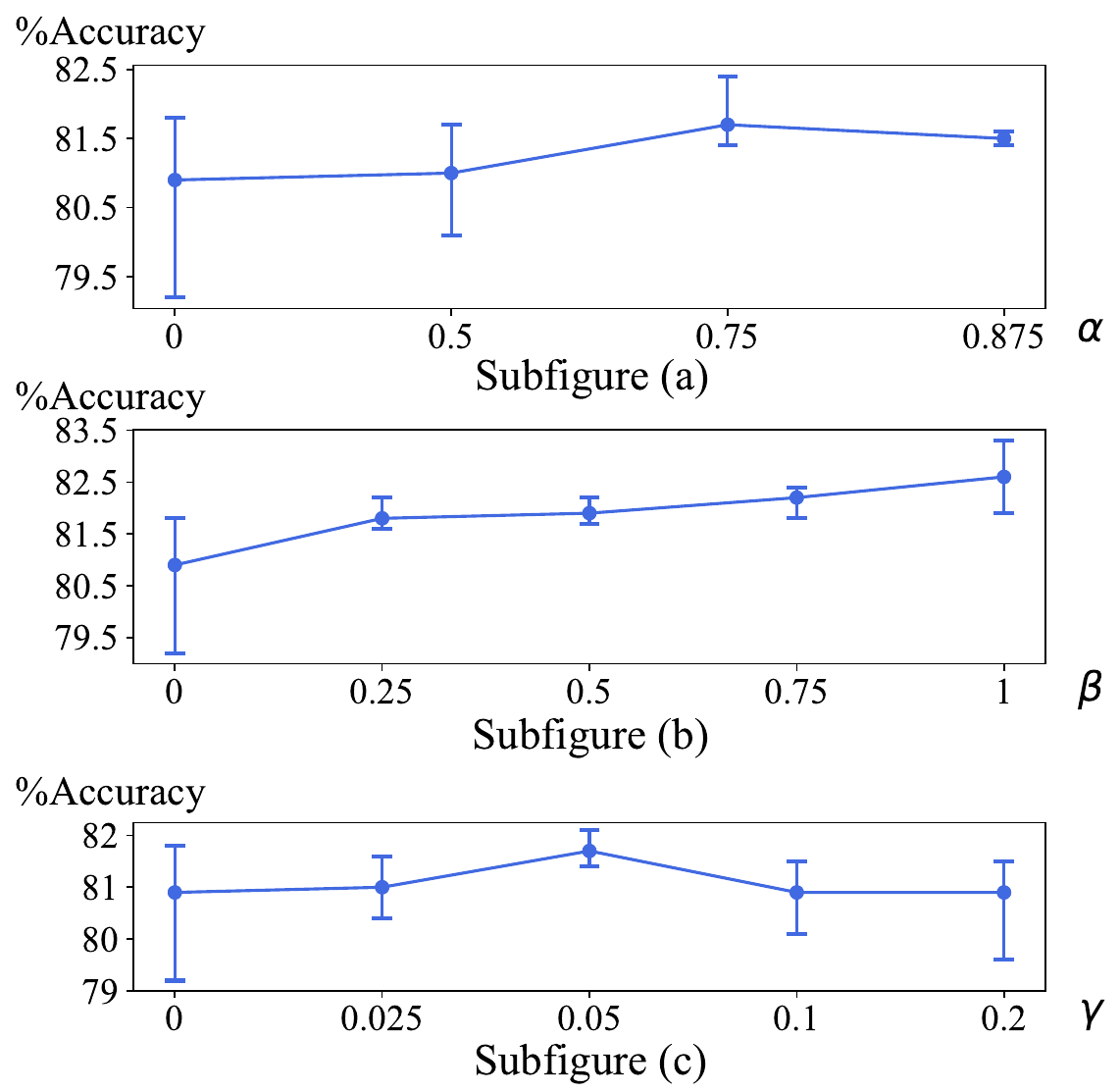}
    \vspace{-0.3cm}
    \caption{The results are all on Leave-one-speaker-out setup. The mean, maximum and minimum accuracies of three results from different random seeds are shown. Subfigures (a), (b), and (c) respectively depict the changes in ASAD accuracy of SW$_\text{CNN}$ influenced by (a) overlapping ratio $\alpha$, (b) time masking ratio $\beta$, and (c) auxiliary loss weighting factor $\gamma$.}
    \vspace{-0.3cm}
    \label{fig:overlap mask multitask}
\end{figure}

Fig.~\ref{fig:overlap mask multitask}(a) depicts the influence of overlapping-based data augmentation, without using time masking or multitask training. The overlapping ratio $\alpha$ is the ratio of the overlapping region length to the decision window length. It is observed that decoding accuracy improves with an increase in $\alpha$ to 0.75, which benefits from the growth of the amount of training data. Further increase $\alpha$ to 0.875 results in a decrease in accuracy, despite the amount of training data being doubled compared to that at a $\alpha=0.75$, indicating that SW$_\text{CNN}$ could suffer from excessive repetition of information.

Fig.~\ref{fig:overlap mask multitask}(b) shows the influence of using time masking-based data augmentation alone. 
The time masking ratio $\beta=\tau / T$ is defined as the ratio of the maximum length of the masked region to the decision window length. It is observed that increasing $\beta$ improves ASAD accuracy, meaning that increasing data diversity through time masking enhances the model's performance.

\subsection{Model Combination of SW$_\text{CNN}$}

In this section, a setup with hyperparameters $\alpha=0.75$, $\beta=1$, and $\gamma=0.05$ was used as our standard-setting thereafter. Nine channels (`Fpz', `Fp1', `AF3', `F5', `Fp2', `AF4', `F6', `AF7', `AF8') located near the eyes are selected and only their data is used to train SW$_\text{CNN}$ with standard setting, denoted as \(M_{\text{NineChan}}\). SW$_\text{CNN}$ trained on all channels with the standard setting is denoted as \(M_{\text{AllChan}}\). The final output is calculated as \(y = 0.5 \cdot \text{Softmax}(M_{\text{AllChan}}(x)) + 0.5 \cdot \text{Softmax}(M_{\text{NineChan}}(x))\), and the system is denoted as ``SW$_\text{CNN}$ combined''.
All results of the three experimental setups are shown in Table~\ref{tab: result}. By combining models from the two distinct configurations, a final accuracy of 84.9\% in the Leave-one-speaker-out setup is achieved, representing a 4.9\% improvement over the previous SOTA benchmark. New SOTA benchmarks are established across all three setups.

\begin{table}[]
\centering
\caption{Summary of results in the literature. The reported results for the CNN\cite{vandecappelle2021eeg} are our replications, as only the median accuracy was provided in the original paper, not the mean. The experimental setups used for DenseNet-3D\cite{xu2023densenet} and EEG-Graph Net\cite{cai2023brain} are similar to ours but not identical. The standard setting with $\alpha=0.75$, $\beta=1.0$ and $\gamma=0.05$ is used for the single SW$_\text{CNN}$ system.}

\begin{tabular}{@{}llc@{}}
\toprule
Experimental Setup                     & Models                                          & \%Accuracy                          \\ \midrule
\multirow{4}{*}{Leave-one-speaker-out} & CNN\cite{vandecappelle2021eeg} & 75.2                           \\
                                       & CSP\cite{geirnaert2020fast}          & 80.0                      \\
                                       & SW$_\text{CNN}$     & 83.2             \\
                                       & SW$_\text{CNN}$ combined     & \textbf{84.9} \\ \midrule
\multirow{3}{*}{Every-trial}           & DenseNet-3D\cite{xu2023densenet}  & 94.3                           \\
                                       & EEG-Graph Net\cite{cai2023brain}    & 96.2                           \\
                                       & SW$_\text{CNN}$   & \textbf{96.9} \\ \midrule
\multirow{3}{*}{Leave-one-subject-out} & CNN\cite{vandecappelle2021eeg} & 65.3           \\
                                       & CSP\cite{geirnaert2020fast} & 66.0        \\
                                       & SW$_\text{CNN}$ & \textbf{71.2} \\ \bottomrule
\end{tabular}

\label{tab: result}
\end{table}

\subsection{SWIM}

SWIM loads the parameters of SW$_\text{CNN}$ trained in a standard setting, while other parameters are randomly initialized. When training SWIM, the total window length is set to 5s, with the last second as the current window and the rest as the previous windows. SWIM needs to determine the auditory attention direction of the subject in the last second. SW$_\text{CNN}$ slides over the 5s window with a window length of 1s and a step size of 0.125s. For data augmentation, the overlapping ratio is set to 0.75, and the time masking ratio is set to 1.0. For comparison, Mamba in SWIM is replaced with a Transformer, and both have approximately the same number of parameters. This modified model is referred to as SWIT, with all other settings remaining the same.  Additionally, a standard-setting trained SW$_\text{CNN}$ is used for testing alongside SWIM and SWIT.

\begin{figure}
    \centering
    \includegraphics[width=1\linewidth]{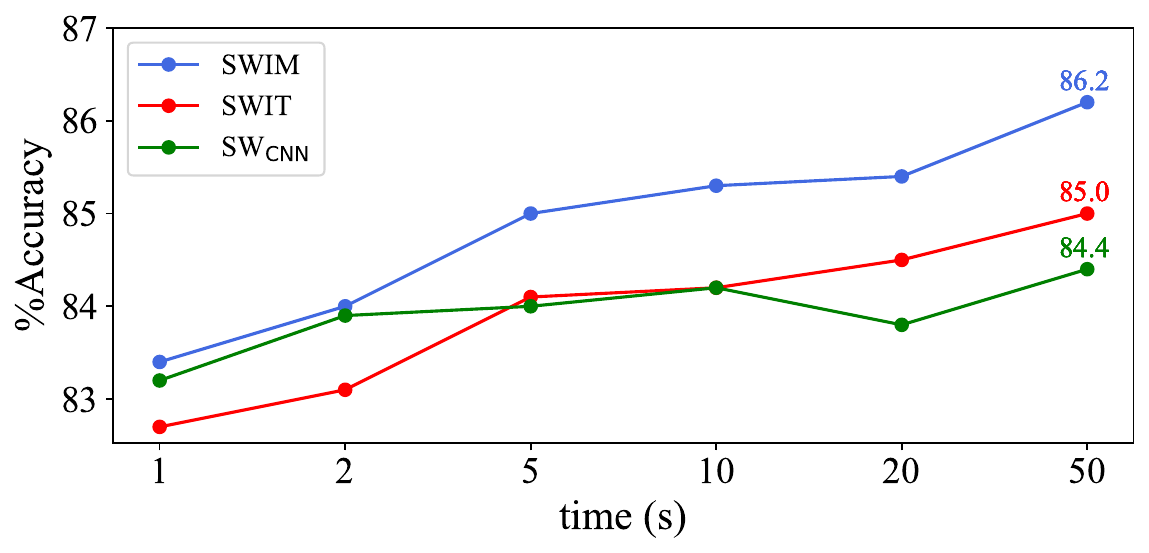}
    \vspace{-0.3cm}
    \caption{The results of SWIM, SWIT and SW$_\text{CNN}$ in Leave-one-speaker-out setup. SWIT refers to the model obtained by replacing Mamba in SWIM with a Transformer. The $x$-axis is the window length the model could use during the test. SWIM achieves the highest accuracy 86.2\% when the window length is 50s, while the accuracy of SWIT and SW$_\text{CNN}$ is 85.0\% and 84.4\% respectively.}
    \vspace{-0.3cm}
    \label{fig:mamba-result}
\end{figure}

The accuracy of SWIM, SWIT and SW$_\text{CNN}$ on different window lengths is shown in Fig.~\ref{fig:mamba-result}. It is observed that SWIM generally outperforms SWIT and SW$_\text{CNN}$ under all window lengths. Let \(t\) represent the window length the model can use during testing. When  \(t = 1s\), although SWIM can not use any extra-temporal information compared to SW$_\text{CNN}$, its accuracy is still slightly higher than that of SW$_\text{CNN}$. As \(t\) increases, the accuracy gradually improves. When \(t=50s\), SWIM can access the most history information, achieving the highest accuracy 86.2\%. This indicates that long-term temporal information can help the classification of auditory attention direction in the current window, suggesting that features in the current window are related to the information from the previous decision windows.
However, in the current KUL dataset, the subjects' attention direction is often fixed. This makes it difficult to distinguish between using history information to determine the attention direction in the current short window and determining the attention direction for the entire window. In the future, we will further test our proposed SWIM on datasets where the attention direction changes usually.

\subsection{Channel Importance of SW$_\text{CNN}$}

\begin{figure}[]
    \centering
    \includegraphics[width=0.6\linewidth]{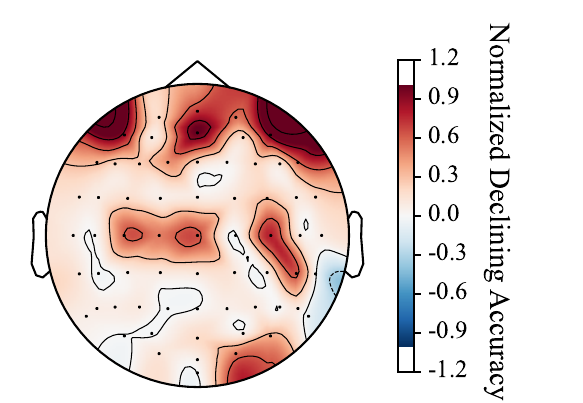}
    \vspace{-0.3cm}
    \caption{EEG topographic map of channel importance, which is normalized declining accuracy when excluding each channel. The redder the channel, the more important it is.}
    \vspace{-0.3cm}
    \label{fig:topo}
\end{figure}

Each of the 64 channels on the test set is sequentially masked, and the ASAD accuracy using SW$_\text{CNN}$ with the standard setting is tested to evaluate the impact of the absence of each channel on accuracy. The importance of each channel is determined by the accuracy difference achieved with all channels and the accuracy obtained when excluding the specific channel.

As depicted in Fig.~\ref{fig:topo}, our findings suggest that channels near the eyes contribute most to successful classification, potentially due to an eye-gaze bias\cite{rotaru2024we}. This bias is characterized by the subject's unconscious tendency to direct their gaze towards the speaker they are attending to. As EEG equipment can inadvertently record these gaze patterns, it allows for the potential exploitation of this gaze information to boost ASAD performance, either intentionally or unintentionally; alternatively, these channels may simply be the most informative, independent of eye-gaze influence. In conclusion, the channels near the eyes should be utilized if they help decode in practice, regardless of whether they contain eye-gaze bias or not. Fig.~\ref{fig:topo} also implies that accuracy improves when certain channels are excluded. This is likely because these channels provide no useful information to detect the auditory attention, but SW$_\text{CNN}$ is marginally over-fitted on them regardless.

\subsection{Model Pattern Analysis}

\begin{figure}
    \centering
    \includegraphics[width=\linewidth]{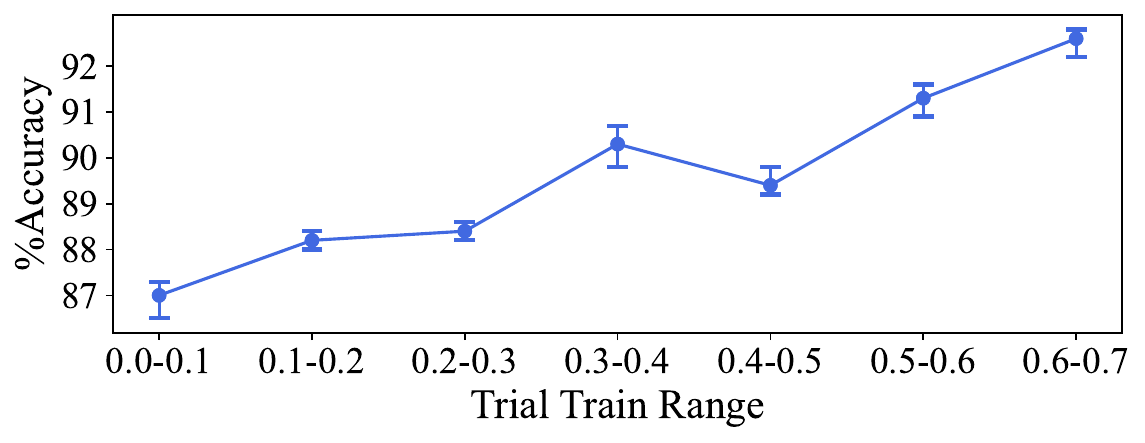}
    \vspace{-0.4cm}
    \caption{The trial train range means the training data is from which range of each trial. For example, the trial train range is 0-0.1 means that the training set is from the first 10\% of each trial.}
    \vspace{-0.3cm}
    \label{fig:trial train}
\end{figure}

Why does the Every-trial setup significantly outperform other setups in terms of accuracy? One hypothesis is that the model learns temporal features due to the consistent attending location label within each trial, a continuous duration \cite{rotaru2024we}. Essentially, the model may identify a test segment as part of a particular trial based on these temporal features and correctly predict the attention location from the corresponding training set. The Every-trial setup's training set includes parts of all trials, facilitating the learning of these temporal features. In contrast, other setups lack this advantage as their training sets exclude trials from the test set, making temporal feature learning challenging.

To verify this assumption, we devised an experiment similar to the Every-trial setup on SW$_\text{CNN}$, but with a variation in the training set. Rather than using the first 70\% of each trial, specific durations, such as 0-10\% or 10-20\%, were used. This approach keeps the amount of training data constant, with the temporal distance between the training and test sets being the sole variable. If our hypothesis holds, accuracy should increase as the training set gets temporally closer to the test set.
Our findings depicted in Fig.~\ref{fig:trial train} largely align with the hypothesis that the model learns temporal features in the Every-trial setup.

\section{Conclusions}


In this paper, SWIM, a short-window CNN integrated with Mamba is proposed. Benefiting from the short window CNN (SW$_\text{CNN}$) and Mamba's long-term sequence modeling ability, SWIM can achieve a highly accurate, efficient and agile response to the user's auditory attention focus. By applying data augmentation strategies, multitask training, and model combination to SW$_\text{CNN}$, an accuracy of 84.9\% in the leave-one-speaker-out setup is achieved, surpassing the previous state-of-the-art benchmark by 4.9\%. SW$_\text{CNN}$ also sets new benchmarks in leave-one-subject-out and every-trial setups. By utilizing long-term historical information, SWIM achieves a final accuracy of 86.2\% in the leave-one-speaker-out setup. Additionally, an analysis of the importance of each channel confirms that channels near the eyes are crucial for decoding accuracy, although it is unclear if this is due to eye-gaze bias. Finally, we confirm that the model becomes over-fitted to the temporal dependency of each trial in the every-trial setup.

\newpage
\bibliographystyle{IEEEbib}
\bibliography{main}

\end{document}